# An iterative tomogravity algorithm for the estimation of network traffic

**Jiangang Fang**[1], **Yehuda Vardi**[1,*] **and Cun-Hui Zhang**[1,†]

*Rutgers University*

**Abstract:** This paper introduces an iterative tomogravity algorithm for the estimation of a network traffic matrix based on one snapshot observation of the link loads in the network. The proposed method does not require complete observation of the total load on individual edge links or proper tuning of a penalty parameter as existing methods do. Numerical results are presented to demonstrate that the iterative tomogravity method controls the estimation error well when the link data is fully observed and produces robust results with moderate amount of missing link data.

## 1. Introduction

This paper concerns the estimation of network traffic based on link data. The traffic matrix of a network, which gives the amount of source-to-destination (SD) flow, is an essential element in a wide range of network administration and engineering applications. However, in today's fast growing communication networks, it is often impractical to directly measure network traffic matrices due to cost, network protocol and/or administrative constraints, while measurements of the total traffic passing through certain individual links are more readily available. Thus, the problem of estimating SD traffic based on link data, called network tomography [10], is of great interest to communications service providers.

In the network tomographic model [10]

$$\mathbf{y} = A\mathbf{x}, \qquad (1.1)$$

where $\mathbf{y}$ is a vector of traffic loads on links, $A = (a_{ij})$ is a known routing matrix with elements $a_{ij} = 1$ if link $i$ is in the path for the $j$-th pair of SD nodes and $a_{ij} = 0$ otherwise, and $\mathbf{x}$ is the SD traffic flow as a vectorization of the traffic matrix. Here the routing protocol $A$ is fixed. In typical network applications, the number of links (edges) is of the same order as the number of nodes (vertices) in the network graph, while the number of SD pairs is of the order of the square of the number of nodes. Thus, $\dim(\mathbf{y}) \gg \dim(\mathbf{x})$ and the network tomographic model (1.1) is ill-posed. Vardi [10] identified the ill-posedness of (1.1) as the main difficulty of network tomography and proposed to estimate the expected traffic flow based on

---

*Research partially supported by National Science Foundation Grant DMS-0405202.
†Research partially supported by National Science Foundation Grants DMS-0405202, DMS-0504387 and DMS-0604571.
[1]Department of Statistics, Hill Center, Busch Campus, Rutgers University, Piscataway, New Jersey 08854, USA, e-mail: alexfang@stat.rutgers.edu; vardi@stat.rutgers.edu; czhang@stat.rutgers.edu

independent copies of **y** by modeling the variance of **y**. The problem has since being considered by many research groups. See Vanderbei and Iannone [9] for MLE/EM, Cao et al. [1, 2] for MLE/EM in the model $x_j \sim N(\lambda_j, \phi\lambda_j^c)$ and non-stationarity issues, Medina et al. [8], Liang and Yu [6] for a more scalable pseudo-likelihood, Liang et al. [7] for additional direct observations of flow data for selected SD pairs, and Coates et al. [4] and Castro et al. [3] for surveys with additional references. In general, these methods require observations of multiple copies of **y**.

An interesting and noticeable development in the area is the introduction of (tomo)gravity algorithms and related methods based on a single snapshot of the network, i.e. one copy of **y**. Zhang et al. [11] observed that in certain communications networks (e.g. a backbone network where each node represents a PoP, or point of presence), almost all the traffic flow is generated by and destined to a known set of edge nodes which do not serve as intermediate nodes in any SD paths. Thus, each SD path begins with a source edge node, traverses through an inbound edge link, an inner network, and then an outbound edge link to a destination edge node. Under this assumption, the total inbound flow $N_s^{(in)}$ from a source node $s$ is the sum of the loads over all the inbound edge links from $s$ and the total outbound flow $N_d^{(out)}$ to a destination node $d$ is the sum of the loads over all the outbound edge links to $d$. The edge nodes communicate to each other through an inner network with a directed graph composed of inner nodes and links and a routing protocol, but the inner nodes does not generate or receive traffic. Moreover, Zhang et al. [11] observed that for each fixed source node $s$, the distribution of the inbound traffic $N_s^{(in)}$ from $s$ to different destinations $d$ is approximately proportional to the total outbound loads $N_d^{(out)}$ these destinations receive. Formally, this is called the gravity model and can be written in Vardi's [10] vectorization as

$$(1.2) \qquad \widetilde{x}_j = \frac{N_{s_j}^{(in)} N_{d_j}^{(out)}}{N}, \quad N = \sum_s N_s^{(in)} = \sum_d N_d^{(out)},$$

where $s_j$ and $d_j$ are respectively the source and destination nodes for the $j$-th SD pair, $\widetilde{x}_j$ is the corresponding component of the simple gravity solution $\widetilde{\mathbf{x}}$ as an approximation of the vector **x** in (1.1), and $N$ is the total flow. The gravity model is best described as

$$(1.3) \qquad \widetilde{x}_{sd} = N_s^{(in)} N_d^{(out)}/N$$

with a slight abuse of notation, where $\widetilde{x}_{sd}$ is the traffic flow from source $s$ to destination $d$ in the gravity model, i.e. $\widetilde{x}_j = \widetilde{x}_{s_j d_j}$. Here, the relationship between the link data **y** and the SD traffic flow **x** is still governed by the tomographic model (1.1). Due to the additional information provided in the gravity model (1.2) about the nature of the SD traffic **x**, the number of unknowns in **x** is square rooted. Thus, the ill-posedness of (1.1) is greatly alleviated. In particular, if all link loads are observed, the total inbound flow $N_s^{(in)}$ and outbound flow $N_d^{(out)}$ for individual edge nodes and thus the total traffic $N$ are all available network statistics in the gravity model. In addition to the gravity solution $\widetilde{\mathbf{x}}$ in (1.2), Zhang et al. [11] developed the simple tomogravity solution

$$(1.4) \qquad \arg\min_{\mathbf{x}} \left\{ \|\mathbf{x} - \widetilde{\mathbf{x}}\| : A\mathbf{x} = \mathbf{y} \right\}$$

and more general tomogravity solutions when the edge nodes are further classified as "access" or "peering", while Zhang et al. [12] developed entropy regularized

tomogravity solution as

$$(1.5) \qquad \arg\min_{\mathbf{x}} \left\{ \|\mathbf{y} - A\mathbf{x}\|^2 + \phi N^2 K(\mathbf{x}/N, \widetilde{\mathbf{x}}/N) \right\},$$

where $K(\cdot, \cdot)$ is the Kullback-Leibler information and $\phi$ is a tuning parameter for the penalty level. These tomogravity solutions require complete knowledge of the total inbound and outbound flow, i.e. $N_s^{(in)}$ and $N_d^{(out)}$, for all individual source and destination nodes. They perform reasonably well when such information is available and have been implemented in certain AT&T commercial networks.

In this paper, we propose an iterative tomogravity (ITG) algorithm which alternately seeks estimates as local optimal solutions in the tomographic space (1.1) and a gravity space of network traffic flow $\mathbf{x}$. Our algorithm, described in Section 2, is based on a single snapshot of the link data and does not require the full knowledge of the total inbound and outbound flow for all individual edge nodes. The idea is to use the gravity space, instead of the specific simple gravity solution (1.2), to regularize the network tomography problem (1.1). In Section 3, we present the results of a real-data experiment to demonstrate that the ITG method is competitive compared with other tomogravity algorithms when the complete link data is available and robust when a moderate amount of link data is missing.

## 2. An iterative tomogravity algorithm

In a general network tomographic model, the observed link data, as a sub-vector $y^*$ of the vector $\mathbf{y}$ in (1.1), satisfies

$$(2.1) \qquad \mathbf{y}^* = A^* \mathbf{x},$$

where the matrix $A^* = (a_{ij}^*)$ is composed of the rows of the routing matrix $A$ in (1.1) corresponding to the observed links, and $\mathbf{x}$ is the SD traffic flow as in (1.1). Let $J$ be the total number of SD-pairs of concern. For the observation $y^*$, the tomographic space of probability vectors is

$$(2.2) \qquad \mathcal{T}^* = \left\{ \mathbf{f} \in \mathbb{R}^J : \mathbf{y}^* \propto A^* \mathbf{f}, \ \mathbf{f} \geq 0, \ \mathbf{1}^T \mathbf{f} = 1 \right\},$$

where $\mathbf{1}$ is the vector composed of 1's and $\mathbf{v}^T$ denotes the transpose of a vector $\mathbf{v}$. Here and in the sequel, inequalities are applied to all components of vectors.

In the literature, different types of flow and load are often specifically denoted. Let $\mathbf{y}^{(\text{net})}$ be the link loads of the inner network, $\mathbf{y}^{(\text{edge})}$ the loads on the links between the edge nodes and inner network, $\mathbf{y}^{(\text{self})}$ the load on the links from the edge nodes to themselves, $\mathbf{x}^{(\text{net})}$ the traffic flow between distinct edge nodes (necessarily through the inner network), and $\mathbf{x}^{(\text{self})}$ the flow of the edge nodes to themselves. Since $\mathbf{x}^{(\text{self})}$ does not go through the inner network and the flow from an edge node to itself is the same as the load on the corresponding self-link, the tomographic model can be written as

$$(2.3) \qquad \mathbf{y} = \begin{pmatrix} \mathbf{y}^{(\text{net})} \\ \mathbf{y}^{(\text{edge})} \\ \mathbf{y}^{(\text{self})} \end{pmatrix} = \begin{pmatrix} A^{(\text{net})} & 0 \\ A^{(\text{edge})} & 0 \\ 0 & I^{(\text{self})} \end{pmatrix} \begin{pmatrix} \mathbf{x}^{(\text{net})} \\ \mathbf{x}^{(\text{self})} \end{pmatrix} = A\mathbf{x},$$

with $I^{(\text{self})}$ being the identity matrix giving $\mathbf{y}^{(\text{self})} = \mathbf{x}^{(\text{self})}$, provided that the inner network does not generate traffic. This is a special case of Vardi's [10] tomographic model (1.1) describing decompositions of the SD traffic **x** and link load **y**, but (1.1) can be also viewed as $\mathbf{y}^{(\text{net})} = A^{(\text{net})}\mathbf{x}^{(\text{net})}$. In this paper, the observed $\mathbf{y}^*$ in (2.1) is a general sub-vector of the **y** in (2.3) to allow partial observation of $\mathbf{y}^{(\text{edge})}$ and networks without $\mathbf{y}^{(\text{self})}$ and $\mathbf{x}^{(\text{self})}$.

Suppose throughout the sequel that the list of the SD-pairs $(s_j, d_j), i = 1, \ldots, J$, forms a product set composed of all the pairings from a set $S$ of source nodes to a set $D$ of destination nodes ($D \neq S$ allowed), so that $J = |S||D|$, where $|C|$ is the size of a set $C$. This gives a one-to-one mapping between $\mathbb{R}^J$ and the space of all $|S| \times |D|$ matrices:

$$\mathbf{v} = (v_1, \ldots, v_J)^T \sim (v_{sd})_{|S| \times |D|}, \quad v_j = v_{s_j d_j}.$$

In this notation, the gravity space of probability vectors is

$$(2.4) \quad \mathcal{G} = \Big\{ \mathbf{g} \in \mathbb{R}^J : \mathbf{g} \sim (g_{sd})_{|S| \times |D|} = \mathbf{p}\mathbf{q}^T, \ \mathbf{g} \geq 0, \ \mathbf{1}^T \mathbf{g} = 1 \Big\},$$

i.e. $g_{sd} = p_s q_d$ or matrices of rank 1, where $\mathbf{p} \in \mathbb{R}^{|S|}$ and $\mathbf{q} \in \mathbb{R}^{|D|}$.

Zhang et al. [11] proposed (1.2) as the simple gravity algorithm and (1.4) as the simple tomogravity algorithm. Zhang et al. [12] proposed (1.5) as the entropy-regularized tomogravity algorithm. Their basic ideas can be summarized as follows: (i) The gravity model gives a rough approximation of the SD flow; (ii) When the simple gravity solution (1.2) is available, it can be used to regularize Vardi's tomographic model (1.1). Motivated by their work, we propose the following algorithm which provides estimates of the SD flow **x** in (2.1).

**Iterative tomogravity algorithm (ITG):**

$$(2.5) \qquad \text{Initialization: } \mathbf{g} = \mathbf{1}/J$$

$$(2.6) \qquad \text{Iteration: } \mathbf{f}^{(\text{new})} = \arg\min \Big\{ K(\mathbf{f}, \mathbf{g}^{(\text{old})}) : \mathbf{f} \in \mathcal{T}^* \Big\}$$

$$(2.7) \qquad \mathbf{g}^{(\text{new})} = \arg\min \Big\{ K(\mathbf{f}^{(\text{new})}, \mathbf{g}) : \mathbf{g} \in \mathcal{G} \Big\}$$

$$(2.8) \qquad \text{Finalization: } \widehat{N} = (\mathbf{1}^T \mathbf{y}^*)/(\mathbf{1}^T A^* \mathbf{f}^{(\text{fin})})$$

$$(2.9) \qquad \widehat{\mathbf{x}} = \widehat{N} \mathbf{f}^{(\text{fin})}$$

where $K(\mathbf{f}, \mathbf{g})$ is the Kullback-Leibler information defined as

$$(2.10) \qquad K(\mathbf{f}, \mathbf{g}) = \sum_{j=1}^J f_j \log \frac{f_j}{g_j}.$$

As mentioned earlier, our basic idea is to use the gravity space (2.4), instead of the simple gravity solution (1.2), to regularize the tomographic model (2.2). A main advantage of this approach is that it does not require the knowledge of the simple gravity solution or equivalently, the complete observation of loads on all edge links. Numerical results in Section 3 demonstrate that when the complete link data **y** is observed, the ITG (2.9) and the entropy-regularized tomogravity (1.5) perform comparably in terms of estimation error, and they both outperform the simple gravity (1.2) and tomogravity (1.4). Moreover, the ITG without using the knowledge of the "access" or "peering" status of links has similar performance compared with the generalized tomogravity method [11] which requires such knowledge. We note that the ITG method does not need a tuning parameter as (1.5) does.

A main difference between ITG (2.9) and the simple tomogravity (1.4) is that the simple gravity solution $\widetilde{\mathbf{x}}$ in (1.2) is not explicitly used in ITG, since $\mathbf{g}$ is treated as an unknown in the ITG algorithm. However, the information in the observed portions of $\mathbf{y}^{(\text{edge})}$ and $\mathbf{y}^{(\text{self})}$ is still utilized in the ITG iterations through the tomographic space (2.2), instead of directly computing $\widetilde{\mathbf{x}}$ from $\mathbf{y}^{(\text{edge})}$ and $\mathbf{y}^{(\text{self})}$ as in (1.3). If the simple gravity $\widetilde{\mathbf{x}}$ (or an approximation of it if $\widetilde{\mathbf{x}}$ is not fully available) is used as the initialization for ITG, the simple tomogravity solution is the result of a single ITG iteration. We may also treat $\mathbf{g} = \widetilde{\mathbf{x}}/N$ as an unknown in (1.5), cf. Section 4, but then a tuning parameter is still required.

We use the relaxation algorithm of Krupp (1979) to compute (2.6) of the ITG, while (2.7) is explicit with

$$g_{sd}^{(\text{new})} = \sum_{d'} f_{sd'}^{(\text{new})} \sum_{s'} f_{s'd}^{(\text{new})}$$

as in (1.3). Here is a full description of the relaxation algorithm. Let $\mathbf{g}^{(\text{old})} = (g_1^{(\text{old})}, \ldots, g_J^{(\text{old})})^T$ be a given probability vector. The problem is to minimize $K(\mathbf{f}, \mathbf{g}^{(\text{old})})$ under the linear constraints in (2.2). Since $y_i^* = 0$ implies $f_{ij} = 0$ for all $j$ with $a_{ij}^* = 1$ and thus reduces the optimization problem to a subset of $j$, we assume $\mathbf{y}^* = (y_1^*, \ldots, y_r^*) > 0$ where $r$ is the total number of links with observed load. Define

$$h_{ij} = \begin{cases} a_{ij}^*/y_i^* - a_{rj}^*/y_r^*, & i = 1, \ldots, r-1, \\ 1, & i = r. \end{cases}$$

The linear constraints $A^* \mathbf{f} = \mathbf{y}^*$ and $\mathbf{1}^T \mathbf{f} = 1$ for the tomographic space (2.2) can be written as $H\mathbf{f} = (\mathbf{0}^T, 1)^T$, where $H = (h_{ij})$. Krupp's [5] relaxation algorithm maximizes

(2.11) $$v_r - \sum_{j=1}^{J} g_j^{(\text{old})} \exp\left\{\sum_{i=1}^{r} h_{ij} v_i - 1\right\}$$

over all vectors $\mathbf{v} = (v_1, \ldots, v_r)^T$ and then set

(2.12) $$f_j^{(\text{new})} = g_j^{(\text{old})} \exp\left\{\sum_{i=1}^{r} h_{ij} v_i - 1\right\}.$$

As (2.11) is concave in $\mathbf{v}$, its optimization is done by the Newton-Raphson method for individual components $v_i$, cycling through $i = 1, \ldots, r$. Since $h_{r,j} = 1$ for all $j$, $\mathbf{f}^{(\text{new})}$ in (2.12) is properly normalized.

The iteration steps (2.6) and (2.7) are both monotone in $K(\mathbf{f}, \mathbf{g})$, so that the ITG algorithm reaches a local minimum of the Kullback-Leibler information between the tomographic (2.2) and gravity (2.4) spaces. However, since $K(\mathbf{f}, \mathbf{g})$ is not convex jointly in $(\mathbf{f}, \mathbf{g})$ with $\mathbf{g}$ in the gravity space, ITG is not guaranteed to converge to a global minimum.

## 3. An example

We conduct numerical experiments with data collected over the Abilene Network (an Internet2 high-performance backbone network in United States) illustrated in

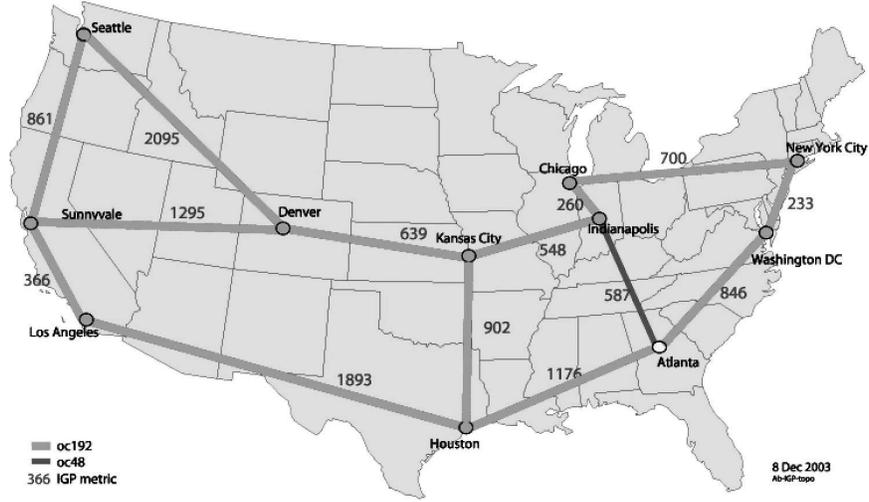

Fig 1. *Abline Network.*

Figure 1, with 12 nodes, 144 total traffic pairs (132 SD pairs and 12 self pairs), 30 inner links, and 24 edge links. We collect the full $12 \times 12$ SD traffic matrices in 5 min intervals for consecutive 19 weeks in 2004. We randomly pick four different periods of 3 days and use the data in these four time periods. We call these four raw datasets as X1, X2, X3, and X4. It turns out that the four datasets give different traffic patterns as the time periods cover different days of the week, cf. Figure 2. For each dataset and each hour, we compute $\mathbf{x}$ as the hourly total SD flow and $\mathbf{y} = A\mathbf{x}$ with a fixed routing matrix $A$ used in the Abilene data.

We compare four procedures using the complete data $\mathbf{y}$ as $\mathbf{y}^*$: the ITG (2.9), the simple tomogravity (STG) in (1.4), the generalized tomogravity (GTG) of Zhang et al. [11] utilizing the extra information of "access" or "peering" status of links, and the entropy regularized tomogravity (ERTG) in (1.5). Since the traffic flow for self pairs $(s = d)$ is directly observable as the load on the self links, the ITG and STG estimate these components of $\mathbf{x}$ without error. Thus, we measure the performance of all estimators by the relative total error for non-self SD pairs

$$(3.1) \qquad \sum_{s \neq d} \left| \widehat{x}_{sd} - x_{sd} \right| \Big/ \sum_{s \neq d} x_{sd},$$

where $x_{sd}$ is the flow from source $s$ to destination $d$. We compute the relative total error for (1.5) with various values of the tuning parameter $\phi$ and found that the performance of (1.5) is near the best in a wide neighborhood of $\phi = 10^{-3} = 0.001$. This confirms the results of Zhang et al. [12]. Thus, $\phi = 10^{-3} = 0.001$ is used for (1.5) in our experiment. We plot the relative total error (3.1) against hour for the four datasets in Figures 3, 4, 5, and 6. We tabulate the average relative error in Table 1. From the results of the experiments, we observed that the performance of the proposed ITG is comparable to the ERTG with the best choice of the tuning parameter and the GTG based on extra information, while all three outperform the STG.

TABLE 1
*Average of relative total errors for 288 different hours (4 different 3-day periods) based on complete link data. The best tuning parameter is used for the ERTG, while extra information is used for the GTG.*

| Method | risk |
|---|---|
| Iterative Tomogravity (ITG) | 0.3001 |
| Entropy Regularized (ERTG) | 0.2995 |
| Simple Tomogravity (STG) | 0.3139 |
| Generalized Tomogravity (GTG) | 0.3026 |

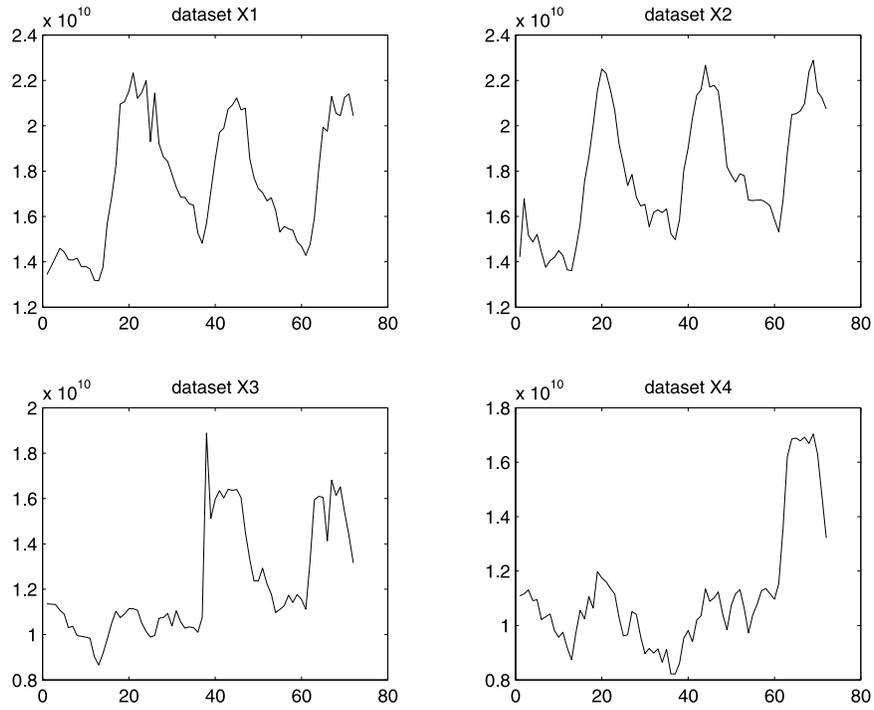

FIG 2. *The total hourly traffic for the 4 non-overlapping 3 day periods.*

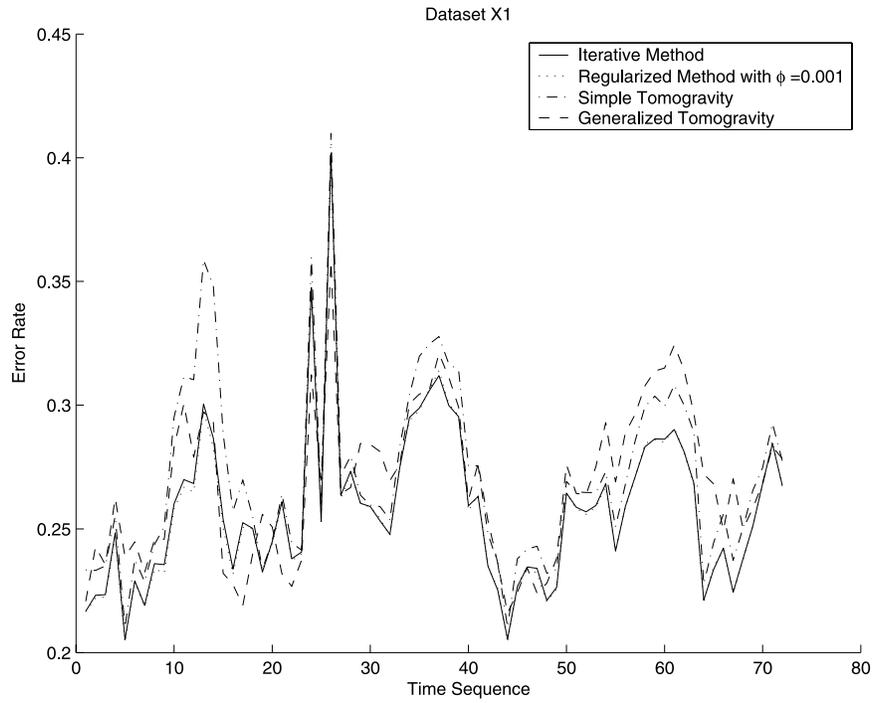

FIG 3. *Compare of the error rate using different models, dataset X1.*

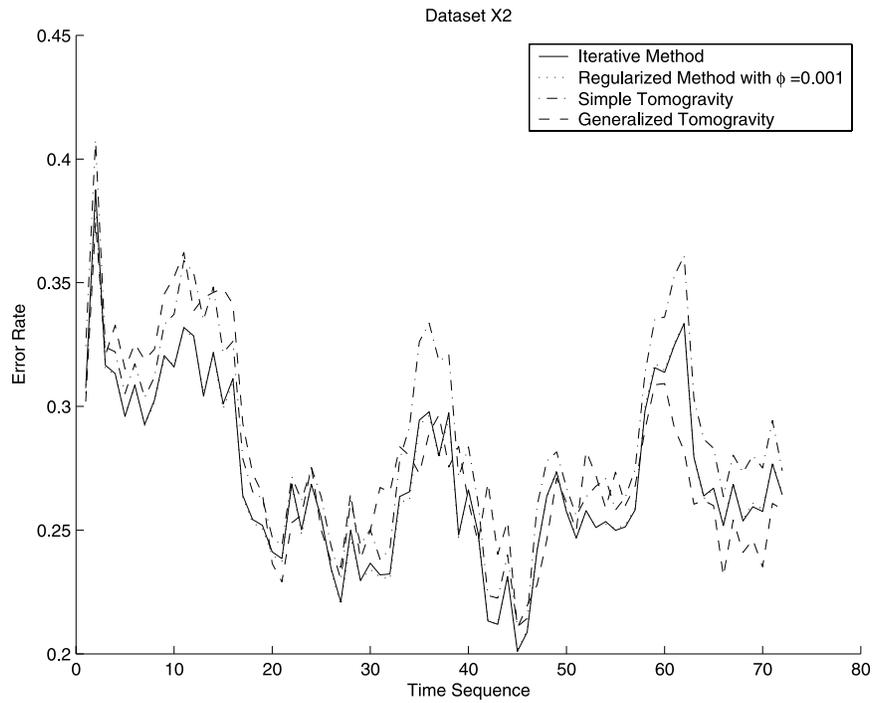

FIG 4. *Compare of the error rate using different models, dataset X2.*

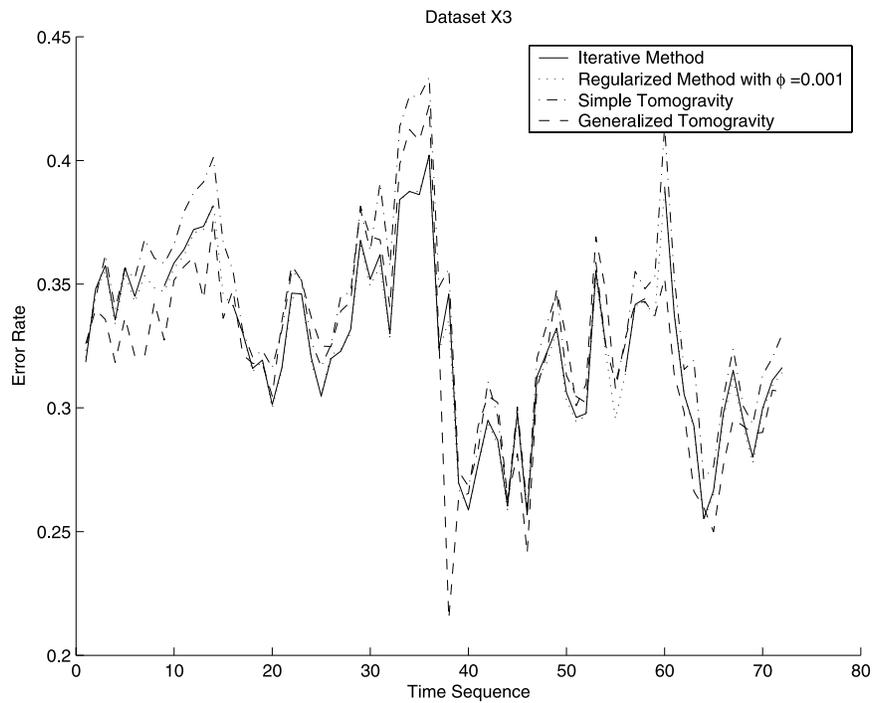

FIG 5. *Compare of the error rate using different models, dataset X3.*

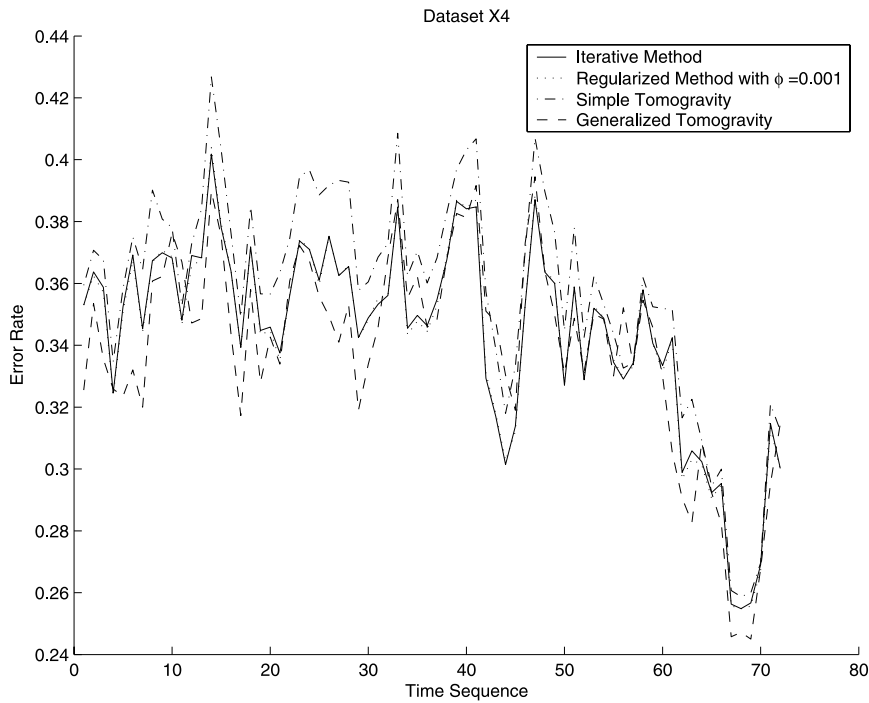

Fig 6. *Compare of the error rate using different models, dataset X4.*

We also exam the relative errors for different SD pairs as functions of the total traffic flow for the SD pairs. We compute the relative total error over 3-day time periods

$$(3.2) \qquad \sum_{t=1}^{t^*} \left| \widehat{x}_{sd}^{(t)} - x_{sd}^{(t)} \right| \Big/ \sum_{t=1}^{t^*} x_{sd}^{(t)}$$

for fixed SD pairs in individual datasets, where $t$ indicates time points with $t^* = 72$. We group the values of (3.2) for SD pairs in all datasets according to the total flow $\sum_{t=1}^{t^*} x_{sd}^{(t)}$ with the grid $\{0, 1/4, 1/2, 3/4, 1, 1.5, 2, 2.5, 3, 4, 5, 7\}$ in the unit of $10^{10}$ packets, and tabulate in Table 2 the average of (3.2) within groups. From Table 2, we observe that the estimation error is essentially a decreasing function of the amount of traffic for individual SD pairs.

Finally, we check the robustness of the ITG (2.9) with missing link data (i.e. $\mathbf{y}^*$ is a proper sub-vector of $\mathbf{y}$). We focus on the case of missing data in edge links as the ITG is the only procedure among the four that do not require observations for all edge links. Let $k$ be the number of edge links with missing data. We use only data for the first day in dataset X1 and compute the average of the relative total error for 10 random missing patterns for each given $k$. We plot this average against $k$ in Figure 7. From Figure 7, we find that the performance of the ITG method is robust against small or moderate amount of missing link data (up to 5 out of 24 edge links).

TABLE 2
*Relative total errors over 72 hours for fixed SD pairs and 3-day periods, grouped according to the total flow. The relative total errors are decreasing functions of the flow for all 4 procedures.*

| Flow Level | # in Group | ITG | ERTG | STG | GTG |
|---|---|---|---|---|---|
| $0 - 0.25$ | 215 | 4.4799 | 5.8725 | 4.5833 | 5.3545 |
| $0.25 - 0.5$ | 100 | 0.4320 | 0.4279 | 0.4548 | 0.4158 |
| $0.5 - 0.75$ | 73 | 0.3457 | 0.3449 | 0.3666 | 0.3467 |
| $0.75 - 1$ | 30 | 0.2997 | 0.2992 | 0.3379 | 0.2505 |
| $1 - 1.5$ | 46 | 0.2286 | 0.2305 | 0.2402 | 0.2588 |
| $1.5 - 2$ | 25 | 0.2878 | 0.2859 | 0.2934 | 0.3089 |
| $2 - 2.5$ | 18 | 0.1836 | 0.1828 | 0.1802 | 0.2080 |
| $2.5 - 3$ | 6 | 0.1583 | 0.1576 | 0.1689 | 0.1207 |
| $3 - 4$ | 7 | 0.1143 | 0.1126 | 0.1335 | 0.1261 |
| $4 - 5$ | 6 | 0.1456 | 0.1448 | 0.1514 | 0.1373 |
| $5 - 7$ | 2 | 0.0887 | 0.0938 | 0.0767 | 0.0882 |

## 4. Discussion

We consider the estimation of SD traffic flow in a network based on observations of a snapshot of traffic loads on links. Based on the ideas of Vardi [10] and Zhang et al. [11, 12], we propose an iterative tomogravity method which allows incomplete observation of the link data. Our main idea is to use the gravity space (2.4), instead of the simple gravity solution (1.2), to regularize Vardi's [10] tomographic model (1.1). A numerical study with a real-life dataset demonstrates that the proposed method has similar performance compared with the methods proposed in [11, 12] which demand complete observation of the link data. We discuss below a number of related issues.

There are two other possible ways of using the gravity space (2.4) to regularize (1.1) that we do not explore in this paper. The first is to use the ITG (2.9) instead of the simple gravity (1.2) in the penalty function in (1.5), resulting in

$$(4.1) \qquad \arg\min_{\mathbf{x}} \left\{ \|\mathbf{y}^* - A^*\mathbf{x}\|^2 + \phi \widehat{N}^2 K\left(\mathbf{x}/\widehat{N}, \mathbf{g}^{(\text{fin})}/\widehat{N}\right) \right\}$$

with the $\widehat{N}$ in (2.8). The second is to alternate between the optimization in the gravity space and entropy-regularized solution, i.e. to replace (2.6) with

$$(4.2) \qquad N^{(\text{new})} = \left(\mathbf{1}^T \mathbf{y}^*\right)/\left(\mathbf{1}^T A^* \mathbf{g}^{(\text{old})}\right)$$

$$\mathbf{f}^{(\text{new})} = \arg\min \left\{ \|\mathbf{y}^* - N^{(\text{new})} A^* \mathbf{f}\|^2 \right.$$

$$(4.3) \qquad \left. + \phi \{N^{(\text{new})}\}^2 K\left(\mathbf{f}, \mathbf{g}^{(\text{old})}\right) : \mathbf{f} \in \mathcal{T}^* \right\}.$$

A small numerical study seems to indicate that there is little difference between (4.1) and the ITG.

The proposed ITG (2.9) implicitly assumes that the measurement error in the tomographic model (2.1) is of smaller order than the bias representing the Kullback-Leibler distance $K(\mathbf{x}/N, \mathcal{G}^*)$ between $\mathbf{x}/N$ and the gravity space (2.4). This seems to be the case in our real-data experiments since ITG significantly improves upon the simple tomogravity (1.4) by formally reducing $K(\mathbf{x}/N, \widetilde{\mathbf{x}}/N)$ to $K(\mathbf{x}/N, \mathcal{G}^*)$. In cases where the measurement error in the tomographic model is potentially of larger order than $K(\mathbf{x}/N, \mathcal{G}^*)$ [or $K(\mathbf{x}/N, \widetilde{\mathbf{x}}/N)$] it would make sense to replace (2.6) by (4.2) and (4.3) in ITG [or to use (1.5)] with a proper tuning parameter $\phi$.

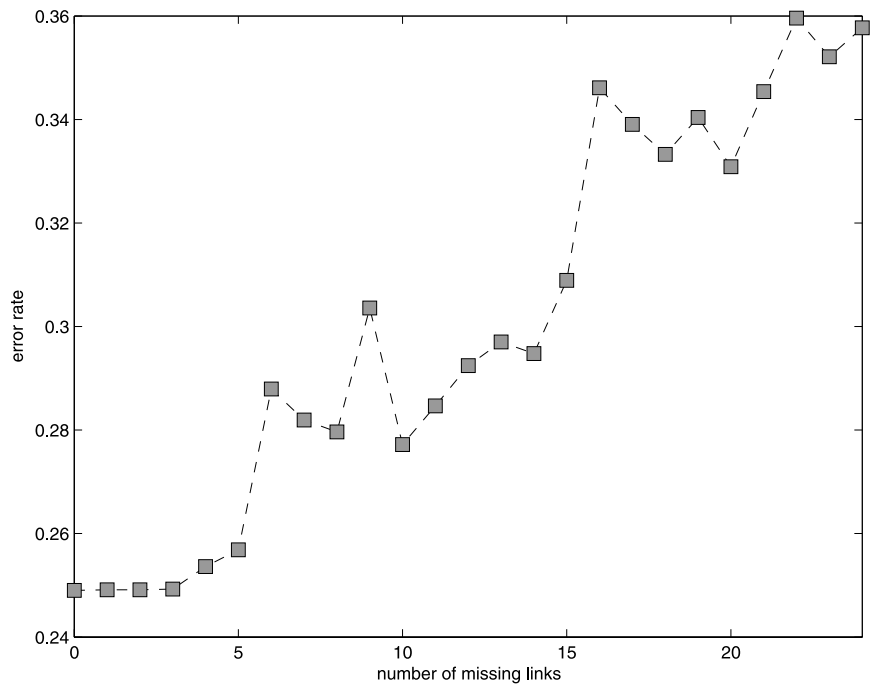

FIG 7. *Relative total errors of the ITG versus the number of edge links with missing data. Average over 10 random missing patterns is used for each point in the plot. The ITG is robust against small or moderate amount of missing link data.*

A possibility to further reduce the bias is to consider the mixed gravity model

$$(4.4) \qquad \mathcal{F}_{\mathrm{mix}} = \left\{ \mathbf{f} : \mathbf{f} = \sum_{k=1}^{k^*} \pi_k \mathbf{f}^{(k)}, \mathbf{f}^{(k)} \in \mathcal{G} \right\}.$$

For example, we may compute a regularized mixed tomogravity solution

$$(4.5) \qquad \arg\min \left\{ \left\| \mathbf{y} - N \sum_{k=1}^{k^*} \pi_k \mathbf{f}^{(k)} \right\| + N^2 \sum_{k=1}^{k^*} \phi_k K(\mathbf{f}^{(k)}, \mathbf{g}^{(k)}) \right\}$$

by alternately optimizing over $\mathbf{g}^{(k)} \in \mathcal{G}$, $\mathbf{f}^{(k)}$, $k = 1, \ldots, k^*$ and the mixing vector $(\pi_1, \ldots, \pi_{k^*})^T$.

It seems that for a network with a fixed routing protocol, the ITG estimate $\widehat{\mathbf{x}}$ in (2.9) is a continuous map of $\mathbf{y}^*$, so that $\widehat{\mathbf{x}} - E\mathbf{x}$ is asymptotically normal when $\mathbf{y}^* - EA^*\mathbf{x}$ is asymptotically normal with $E\mathbf{x}/N \in \mathcal{G}$, as $N \to \infty$. Our simulation study in a small artificial network has demonstrated the validity of this asymptotic normality theorem for moderate sample sizes.

Estimation of traffic matrix based on link-load data alone is difficult as the estimation error is typically above 20%. More accurate results can be obtained if additional information can be extracted from packets passing through routers. See for example Zhao, Kumar, Wang and Xu [13].